\documentclass{article}

  \usepackage{graphicx}

\usepackage{booktabs,tabularx,makecell,caption}

\usepackage{array} 
\PassOptionsToPackage{numbers,compress}{natbib}
\usepackage{natbib} 
 \usepackage[dblblindworkshop, final]{neurips_2025}
 \workshoptitle{The Second Workshop on GenAI for Health: Potential, Trust, and Policy Compliance}

\usepackage[utf8]{inputenc} 
\usepackage[T1]{fontenc}    
\usepackage{hyperref}       
\usepackage{url}            
\usepackage{booktabs}       
\usepackage{amsfonts}       
\usepackage{nicefrac}       
\usepackage{microtype}      
\usepackage{xcolor}         

\title{Application of Whisper in Clinical Practice: the
Post-Stroke Speech Assessment during a Naming
Task}

\author{%
  \makebox[\linewidth][c]{%
    \parbox{0.98\linewidth}{\centering
      Milena Davudova\textsuperscript{1}\thanks{Equal contribution.}\;
      Ziyuan Cai\textsuperscript{1}\footnotemark[1]\;
      Valentina Giunchiglia\textsuperscript{1,2}\;
      Dragos C.\ Gruia\textsuperscript{2}\\
      \textbf{Giulia Sanguedolce}\textsuperscript{2}\;
      \textbf{Adam Hampshire}\textsuperscript{1}\thanks{Co-senior and co-corresponding author.}\;
      \textbf{Fatemeh Geranmayeh}\textsuperscript{2}\footnotemark[2]\\[0.3ex]
      {\normalsize
      \textsuperscript{1}Department of Neuroimaging, King's College London, UK\quad
      \textsuperscript{2}Department of Brain Sciences, Imperial College London, UK\\[-0.2ex]
      \texttt{\{milena.davudova, ziyuan.1.cai, adam.hampshire\}@kcl.ac.uk}\\
      \texttt{\{v.giunchiglia20, dragos-cristian.gruia19, g.sanguedolce22, fatemeh.geranmayeh00\}@imperial.ac.uk}
      }
    }%
  }%
}

\begin{document}

\maketitle

\begin{abstract}
Detailed assessment of language impairment following stroke remains a cognitively complex and clinician-intensive task, limiting timely and scalable diagnosis. Automatic Speech Recognition (ASR) foundation models offer a promising pathway to augment human evaluation, but their effectiveness in the context of speech and language impairment remains uncertain. In this study, we evaluate whether Whisper, a state-of-the-art ASR foundation model, can be applied to transcribe and analyze speech from patients with stroke during a picture-naming task. We assess both verbatim transcription accuracy and the model’s ability to support downstream prediction of language function, which has major implications for outcomes after stroke. Our results show that the baseline Whisper model performs poorly on single-word speech utterances. Nevertheless, fine-tuning Whisper significantly improves transcription accuracy (reducing Word Error Rate by 87.72\% in healthy speech and 71.22\% in speech from patients). Further, learned representations from the model enable accurate prediction of speech quality (average F1 Macro of 0.74 for healthy, 0.75 for patients). However, evaluations on an unseen (TORGO) dataset reveal limited generalizability, highlighting the inability of Whisper to perform zero-shot transcription of single-word utterances on out-of-domain clinical speech and emphasizing the need to adapt models to specific clinical populations. While challenges remain in cross-domain generalization, these findings highlight the potential of foundation models, when appropriately fine-tuned, to advance automated speech assessment and rehabilitation for stroke-related impairments.
\end{abstract}

\section{Introduction}
Stroke is one of the leading causes of adult death and disability worldwide, with global incidence of 12.2 million per year \cite{Feigin_2022}.  One of the most debilitating and common consequences after stroke is aphasia, an acquired disorder of speech and language production and comprehension, with over 30\% prevalence in patients with stroke \cite{Flowers_2016, Hartman_1981, Wade_1986}.  Due to high heterogeneity of symptoms, diagnosis of aphasia requires an in-depth assessment to uncover specific impairments and tailor speech therapy to individual patient’s needs. Computerized assessments provide a cost-effective and accessible platform for post-stroke speech evaluation \cite{Gruia_2023}, but they often depend on manual transcription and scoring by trained clinicians, which is time-consuming, prone to inconsistencies, and difficult to scale.

Advancements in automatic speech recognition (ASR) methods can fill this gap by automating the speech evaluation pipelines. The extensive linguistic knowledge embedded in foundation ASR models, acquired through pretraining on large and diverse datasets, has the potential to be leveraged in a zero and few shot learning context to automate post-stroke speech evaluation pipelines \cite{kumar2022finetuningdistortpretrainedfeatures,Church_2021}. However, whilst the latest ASR models perform exceedingly well on healthy speech data \cite{Radford_2023}, several limitations have hindered the translation of ASR to pathological speech analysis.  Firstly, current ASR models have been trained on healthy speech, limiting their generalizability to pathological speech. Secondly, the potential variability in speech impairments in aphasia necessitates large clinical speech databases for effective training of ASR models, which are currently relatively limited.

In this study, we aimed to assess the performance of Whisper \cite{Radford_2023}, a state-of-the-art transformer-based ASR model pretrained on 680,000 hours of multilingual data, on a stroke-specific speech database derived from a commonly used Naming task. We fine-tuned Whisper on speech from age-matched healthy older adults and patients with stroke and compared the verbatim transcription performance of the fine-tuned and the baseline Whisper models. Further, we evaluated whether Whisper-derived representations could be effectively applied to clinically-relevant downstream tasks, specifically speech impairment severity prediction. Lastly, we assessed the quality of these predictions by conducting a divergent and convergent validity analysis with patients' known clinical features.

\section{Related Work}
Current approaches for automated speech impairment analysis involve three key steps. First, speech-to-text ASR models are employed to generate speech transcriptions from audio recordings. Then, a feature extraction process is applied to the produced transcriptions and audio data. The extracted lexico-acoustic features are used in downstream analyses, such as classification and regression, to estimate the degree of speech impairment \cite{Qin_2020}. Different algorithms have been used for each step.

\subsection{Recurrent Neural Networks}
Previous studies have trained customized ASR models to transcribe aphasic speech, subsequently deriving acoustic and transcription-based features to estimate aphasia severity \cite{le_2018,Qin_2020}. Le and colleagues \cite{le_2018}  utilized a Bidirectional Long-Short Memory-Recurrent Neural Network (BLSTM-RNN) model trained on acoustic features of aphasic speech derived from the AphasiaBank dataset \cite{Macwhinney_2011} and frame-level senone and monophone labels, obtained through a Hidden Markov Model-Gaussian Mixture Model (HMM-GMM) system, to produce speech transcriptions. Similarly, Qin and colleagues \cite{Qin_2020} utilized a Time-Delay Neural Network combined with a Bidirectional Long-Short Term Memory (TDNN-BLSTM) model to transcribe Cantonese aphasic speech. The model was trained on acoustic features derived from healthy speech and HMM-GMM-derived triphone target labels \cite{Qin_2020}. The ASR transcriptions were used to extract features based on text alone \cite{le_2018} or in combination with audio features \cite{Qin_2020} for downstream estimation of aphasia severity by regression. While the above studies reached promising accuracy, the complex architectures of employed ASR systems required a multi-step training process, limiting global optimization \cite{Wang_2019}. Moreover, the models were based on RNN variants, which are less effective than Transformers in modeling long-range dependencies \cite{Bai_2021}. 

\subsection{Transformers}
Transformers have revolutionized the field of ASR by implementing self-attention mechanisms, which capture temporal relationships across the entire input sequence simultaneously, thus overcoming the range limitations of RNN-based architectures and enabling more efficient encoding of contextual information \cite{Vaswani_2023,Mehrish_2023}. Multiple ASR transformer-based models are currently available for speech-to-text transcription including Transformer Transducer \cite{Zhang_2020}, wav2vec 2.0 \cite{Baevski_2020}, SpeechStew \cite{Chan_2021}, Conformer \cite{Gulati_2020} and Whisper \cite{Radford_2023}. However, while transformer-based models demonstrate exceptional performance on healthy speech ASR \cite{Radford_2023}, they often encounter challenges adapting to aphasic speech recognition due to the high heterogeneity in lexico-acoustic features of post-stroke speech.
Recent studies have leveraged pretrained transformer-based ASR models, adapting them to enhance performance on aphasic speech. Within this framework\cite{Torre_2021}, XLSR-53, a model built upon wav2vec 2.0 architecture, was fine-tuned to adapt it to aphasic speech transcription. XLSR-53 \cite{Conneau_2020}, pretrained on the extensive dataset which contains 56,000 hours of unlabeled audio, uses convolutional neural network layers to generate latent speech representations \cite{Baevski_2020}. These representations are then passed through a transformer network to encode contextualized speech features \cite{Baevski_2020}. However, wav2vec 2.0 architecture is outperformed by other transformer-based foundation models such as Whisper \cite{Radford_2023, Yerramreddy_2024}. Whisper-derived representations have been successfully utilized in previous studies to classify dysfluencies in stuttering \cite{Ameer_2024} and to estimate dysarthria severity \cite{Rathod_2023}. Further, Whisper and XLSR-53 model transcripts have been successfully used for relevant speech feature extraction, later used for aphasia type classification, reaching an average F1 score of 90.6 \cite{Wagner_2023}.  Recent work also investigated  hyperparameter tuning and fine-tuning of Whisper to better fit aphasic speech, reaching promising results of 38.5\% and 21.93\% WER on aphasic speech transcription, respectively  \cite{Sanguedolce_2023, sanguedolce2024universal}. However, the vast majority of previous work has focused on sentential speech, and the applications of Whisper to single-word utterances remain to be studied. 

\section{Method}

\subsection{Dataset}
Speech data from age-matched controls and patients with stroke were derived from participants undergoing speech and cognitive testing as part of the the IC3 study \cite{Gruia_2023}. Specifically the data was obtained from the Naming task, modified from the 30-item Boston Naming Task \cite{Roth_2011}, where participants were instructed to name a depicted black and white line drawing picture with a single word. The speech recordings from this task, amongst other speech production tasks, have contributed to a larger speech database SONIVA (Speech Recognition Validation in Aphasia) \cite{Sanguedolce_2025}. The data used in this study will henceforth be referred to as SONIVA-Naming. The accuracy of the Naming task performance was evaluated across each trial using four accuracy metrics: semantic content (Semantic), fluency of articulation (Dysfluency), presence of self-corrections in the response (Self-correction) and phonological correctness (Phonology). Each trial was scored by trained raters on a 3-point scale: 0, 1, or 2, where 2 corresponds to the highest accuracy. The verbatim transcriptions of the audio files ranged from single word to sentences, depending on the participants responses. The inter-rater reliability against a qualified speech therapist was high (Mean Intraclass Correlation Coefficient = 0.83\(\pm\) 0.0046),  and all ambiguous cases were resolved by consulting a speech therapist. In total, the SONIVA-Naming healthy dataset comprised 3960 Naming task trials from 132 healthy participants, while the SONIVA-Naming patient dataset consisted of 2609 trials from 87 patients. Demographic characteristics of the sample are described in Appendix A. There was no group difference for age  (Mann-Whitney U statistic: $5542.50, p = 0.88$). Patients had lower education ($\chi^{2}$=26.19, $p<0.001$) and a higher proportion of non-native English speakers ($\chi^{2}$=4.02, $p=0.04$). A noise-free, single-word, baseline synthetic dataset was additionally generated using the Google Text-to-Speech library for Python \cite{gTTS}. Seven English accents were used to generate a total of 210 synthetic audio files across the 30 stimuli words available in the Naming task.


An unseen dataset was used to assess model generalizability for single-word pathological speech transcription. This dataset was derived from the TORGO database \cite{rudzicz2012torgo}, which includes speech from individuals with dysarthria, a motor speech disorder common in cerebral palsy and amyotrophic lateral sclerosis. Single-word recordings were used to evaluate model transcription of short-form data. The TORGO dataset comprised 1,240 utterances from 7 participants (4 male, 3 female).

\subsection{Model}
OpenAI’s Whisper \cite{Radford_2023} was fine-tuned for the verbatim transcription task. For the accuracy classification task, we employed WhisperForAudioClassification, an adaptation that retains the original Whisper encoder but replaces the decoder with a linear classification head, implemented via the Hugging Face Transformers library \cite{WhisperForAudioClassification}. Whisper architecture is described in Appendix B.


\label{table1}
 
\subsection{Data Pre-processing}
All audio data were resampled to 16 kHz. In the main dataset used for model training and evaluation (SONIVA-Naming), recordings without at least one ground truth accuracy metric (Semantic, Dysfluency, Self-correction, Phonology) were excluded when fine-tuning models for the accuracy score prediction task (n = 23 and n = 114 for SONIVA-Naming healthy and patient datasets, respectively). Data were divided into training, validation, and test sets in a 7:1:2 ratio. Splitting was performed at the participant level such that all trials from the same participants were in the same partition. Due to high class imbalance across accuracy scores, the training sets were randomly downsampled to the minority class in each metric. This resulted in 144, 204, 90, and 21 audio files for the SONIVA-Naming healthy dataset and 198, 729, 108, and 171 trials for the patient dataset, respectively.

\subsection{Model Training}
\paragraph{Verbatim Transcription Task}
Whisper was separately fine-tuned on the training set of synthetic, SONIVA-Naming healthy, and SONIVA-Naming patient data, as well as combined datasets of all participant data (SONIVA-Naming healthy and SONIVA-Naming patient) and all available data (synthetic, SONIVA-Naming healthy and SONIVA-Naming patient) (See Appendix C). Small and Medium Whisper models \cite{openai2024whispersmall, openai2024whispermedium} were used to determine the optimal model size. In total, ten fine-tuned models were obtained: fine-tuned on synthetic data (\textit{ft-syn}), on SONIVA-Naming healthy data (\textit{ft-h}), on SONIVA-Naming patient data (\textit{ft-p}), on all participant data (\textit{ft-hp}) and all available data (\textit{ft-all}), each in Small and Medium sizes.

Models were trained on a single NVIDIA RTX 6000 GPU for a maximum of 1000 steps, with a batch size of 16, which took approximately 2-4 hours. A learning rate of $1 \times 10^{-5}$ was used, with a linear learning rate scheduler incorporating 250 warm-up steps. The model parameters were updated with AdamW optimizer using Cross-Entropy loss function. 


 Models were evaluated on the validation set (batch size of 8) at 50 step intervals. The best model was chosen based on Word Error Rate (WER) on the validation set and was used for final evaluation on the test sets. The evaluation was completed on four different test sets, namely the synthetic, SONIVA-Naming healthy, and SONIVA-Naming patient as well as the unseen TORGO database.\vspace{0.3cm}

\paragraph{Accuracy Prediction Task} The setup was formulated as a multi-class classification task, where one of three possible accuracy scores (0, 1 or 2) was predicted for each given Naming task trial recording in the SONIVA-Naming healthy and patient datasets (See Appendix C).

The encoder was initialized either with the baseline pretrained encoder weights (Whisper without fine-tuning) or with the encoder weights of the models fine-tuned for verbatim transcription of healthy and patient speech. The \textit{ft-h} and \textit{ft-p} model weights were used when predicting accuracy scores on healthy and patient data, respectively, to capture the most relevant speech representations of each group. This setup resulted in a total of 32 models: models were trained separately for each accuracy metric (Semantic, Dysfluency, Self-correction, Phonology), with two encoder weight configurations (fine-tuned or baseline), each in two sizes (Small and Medium) \cite{openai2024whispersmall, openai2024whispermedium} for each dataset separately (SONIVA-Naming healthy and SONIVA-Naming patient). During training, the encoder weights were frozen and only the linear classification head was trained (i.e., linear probing). 

Models were trained on a single NVIDIA RTX 6000 GPU for a maximum of 8000 steps and with a batch size of 16, which took approximately 6-8 hours. A learning rate of $1 \times 10^{-5}$  was used, with a linear learning rate scheduler incorporating 500 warm-up steps. Models were evaluated on the validation set at 100 step intervals. The model parameters were updated with AdamW optimizer using Cross-Entropy loss function. The best model was chosen based on F1 Macro metric on the validation set and was used for final evaluation on the test set. 

\vspace{0.3cm}

\subsection{Evaluation metrics}
\paragraph{Word Error Rate (WER)}
The WER was defined based on the string edit distance, which consists of the ratio of necessary string modifications needed to convert the model prediction into the ground truth label divided by the number of total words spoken (\ref{wer-eq}).

\begin{equation}
  {WER} = \frac{S+I+D}{N}
  \label{wer-eq}
\end{equation}
where S is the number of substitutions, I of insertions, D of deletions, and N is the total number of words spoken. The score was then multiplied by 100 to obtain a percentage measure.\vspace{0.3cm}

\paragraph{F1 Macro}
Performance on the accuracy score prediction task was assessed based on binarized F1 scores. Specifically, the accuracy score 1 was merged with 0 to derive an impaired class (class 0), while accuracy score of 2 was converted to 1 (class 1) to represent unimpaired responses.  The F1 score for the impaired and unimpaired  classes were calculated, and then the average between the two was used to obtain the F1 macro, which was used as the main performance metric (\ref{f1macro}).

\begin{equation}
  {F1_{Macro}} = \frac{\sum_{i=1}^{n} F1_i}{n}
  \label{f1macro}
\end{equation}
where n is the number of classes.

\vspace{0.3cm}

\paragraph{Target Word Detection Accuracy}
Verbatim transcriptions of the SONIVA-Naming dataset trials often contained words beyond the target Naming task word. A target word detection accuracy was calculated by assessing whether the predicted transcriptions correctly identified the target word only. A prediction was classified as a True Positive if the target word appeared in both the ground truth and prediction and a True Negative if absent in both. The presence of the target word in the ground truth label but not in the prediction indicated a False Negative  and the reverse - a False Positive. \vspace{0.3cm}


\subsection{Statistical Analysis}
\noindent

\paragraph{Model size and type analysis} The effect of model type (i.e., baseline vs fine-tuned) and size (i.e., Small vs Medium) on WER across all trials was assessed on the test set using the Friedman test, followed by Mann-Whitney U tests with Bonferroni correction for multiple comparisons. The non-parametric tests were used due to the non-normal distributions of the WER within each dataset. \vspace{0.3cm}

\paragraph{Predictive Validity Analysis}
To evaluate the clinical validity of the predicted speech accuracy scores, an overall predicted accuracy score was derived for each patient in the SONIVA-Naming patient test set ($n=18$) by averaging their predicted trial-by-trial scores for each metric. Patients were classified as either impaired or unimpaired on each predicted accuracy metric. Patients with an overall predicted score smaller or equal to 1 were classified as impaired. Otherwise, they were classified as unimpaired. Impairment status was validated against multiple known clinical and demographic factors. Convergent validity was tested against a manually assessed speech fluency metric, stroke history and English as a second language status – hypothesized to affect predicted impairment status. Divergent validity was tested against sex, low density lipoprotein (LDL) cholesterol levels and smoking status, hypothesized to not show any relationship with predicted impairment status.
\vspace{0.3cm}

To complete the analysis, all categorical variables (e.g., sex, smoking status, English as second language, previous stroke history), were one-hot encoded. The distribution of continuous variables (LDL cholesterol, speech fluency) was tested for normality using the Shapiro-Wilk test. Since the assumption of normality was met, continuous variables between groups were compared with a Student’s t-test. Categorical variables were compared between groups using Fisher’s exact test.  

\section{Results}

\subsection{Verbatim Transcription}

A significant effect of model size and type on WER was detected when testing on synthetic ($\chi^2$(11, $N$=42) = 313.95, $p<0.001$), SONIVA-Naming healthy ($\chi^2$(11, $N$=806) = 5215.60, $p<0.01$), and SONIVA-Naming patient datasets ($\chi^2$(11, $N$=524) = 2442.80, $p<0.01$). All fine-tuned models significantly outperformed baseline Whisper in transcription accuracy on the synthetic, healthy and patient datasets ($p<0.01$) (Table \ref{wer_final_grouped}). Fine-tuning on healthy and patient speech resulted in  87.72\% and 71.22\% improvement in WER for healthy and patient speech compared to baseline model of the same size. A significant improvement was also observed in case of synthetic data, where the best-performing models were Medium \textit{ft-h}, Medium \textit{ft-p} and Small and Medium \textit{ft-all}, reaching a WER of 0\%, compared to a WER of minimum 85.71\% in case of Whisper baseline. The best-performing model on the healthy data was the Small \textit{ft-h}, reaching average WER of 8.82\%. On the patient dataset, Medium \textit{ft-hp} reached the lowest average WER of 26.35\%. However, this Medium model was not significantly better than its Small counterpart ($p>0.05$). Additionally, when evaluated on the patient dataset, the difference in performance between \textit{ft-hp} and single-dataset trained models \textit{ft-h} and \textit{ft-p} was insignificant ($p>0.05$) in both model sizes.

On the unseen TORGO database, a significant effect of model size and type on WER was also detected ($\chi^2$(11, $N$=1240) = 850.08, $p<0.001$) (Table \ref{wer_final_grouped}). All fine-tuned models significantly outperformed baseline Whisper in verbatim transcription accuracy ($p<0.01$). The best performing model was Medium \textit{ft-syn}, which achieved a 22.64\% reduction in WER compared to the baseline model in the same size. However, it did not perform significantly better than other fine-tuned models ($p>0.5$), with the exception of Medium \textit{ft-hp} ($p=0.045$).

\begin{table}[ht]
  \centering
  \caption{Comparison of word error rate (WER) across testing datasets.}
  {\fontsize{8}{9.2}\selectfont 
    \begin{tabular}{@{}%
      l!{\vrule width 0.3pt}%
      l!{\vrule width 0.3pt}%
      c!{\vrule width 0.3pt}%
      c!{\vrule width 0.3pt}%
      c!{\vrule width 0.3pt}%
      c%
    @{}}
    \toprule
    \multicolumn{6}{c}{\textbf{Word error rate (\%)}}\\
    \midrule
    \textbf{Model} & \textbf{Size} & \textbf{Synthetic} & \makecell{\textbf{SONIVA}\\\textbf{healthy}} & \makecell{\textbf{SONIVA}\\\textbf{patient}} & \textbf{TORGO} \\
    \midrule
    Baseline & Small  & 92.85 & 96.54 & 100.54 & 97.82 \\
             & Medium & 85.71 & 90.87 & 97.57  & 96.13 \\
    ft-syn   & Small  & 45.23 & 53.64 & 75.83  & 77.72 \\
             & Medium & 21.42 & 48.06 & 71.53  & \textbf{73.49} \\
    ft-h     & Small  & 2.38  & \textbf{8.82} & 28.38  & 77.39 \\
             & Medium & \textbf{0} & 9.89  & 27.72  & 75.65 \\
    ft-p     & Small  & 7.14  & 14.60 & 28.82  & 75.01 \\
             & Medium & \textbf{0} & 11.94 & 26.80  & 74.62 \\
    ft-hp    & Small  & 2.38  & 12.69 & 27.79  & 78.92 \\
             & Medium & 2.38  & 12.58 & \textbf{26.35} & 79.67 \\
    ft-all   & Small  & \textbf{0} & 12.00 & 27.04  & 78.10 \\
             & Medium & \textbf{0} & 9.07  & 29.15  & 76.58 \\
    \bottomrule
  \end{tabular}
  }
  \label{wer_final_grouped}
\end{table}

\subsection{Target Word Detection accuracy} 
Compared to baseline Whisper, fine-tuned models had a higher target word detection accuracy in both healthy and patient speech (Table \ref{accu}). For healthy speech, the best-performing model was the Small \textit{ft-h}, reaching an accuracy of 0.97. This increase in performance was due to the \textit{ft-h} model making noticeably fewer False Negative mistakes ($n=27$) compared to baseline Whisper ($n=496$). For patient speech, the best-performing models were the Small and Medium \textit{ft-hp}, reaching an accuracy of 0.92. This model made fewer False Negative mistakes ($n=39$) than the Medium baseline model ($n=286$). However, Small and Medium \textit{ft-hp} models produced 2 False Positive mistakes, compared to 1 and 0 False Positive mistakes in Small and Medium baseline models, respectively. Overall, an improvement of up to 61\% and 51\% was observed for healthy and patient data, respectively.

\begin{table}[ht]
  \centering
  \caption{Model target word detection performance on the SONIVA-Naming dataset.}
  {\fontsize{8}{9.2}\selectfont
    \begin{tabular}{
    l!{\vrule width 0.3pt}
    l!{\vrule width 0.3pt}
    l!{\vrule width 0.3pt}
    c!{\vrule width 0.3pt}
    c!{\vrule width 0.3pt}
    c!{\vrule width 0.3pt}
    c!{\vrule width 0.3pt}
    c
  }
    \toprule
    \textbf{Testing Dataset} & \textbf{Size} & \textbf{Model} & \textbf{Accuracy} &
    \makecell{\textbf{True}\\\textbf{Positive}} &
    \makecell{\textbf{True}\\\textbf{Negative}} &
    \makecell{\textbf{False}\\\textbf{Positive}} &
    \makecell{\textbf{False}\\\textbf{Negative}} \\
    \midrule
    SONIVA-Naming healthy & Small  & Baseline & 0.36 & 256 & 24 & 0 & 496 \\
                          &        & ft-syn   & 0.60 & 442 & 24 & 0 & 310 \\
                          &        & ft-h     & \textbf{0.97} & 725 & 24 & 0 & 27 \\
                          &        & ft-p     & 0.92 & 693 & 24 & 0 & 59 \\
                          &        & ft-hp    & 0.96 & 723 & 24 & 0 & 29 \\
                          &        & ft-all   & 0.96 & 719 & 24 & 0 & 33 \\
    \cmidrule(l){2-8}
                          & Medium & Baseline & 0.40 & 284 & 24 & 0 & 468 \\
                          &        & ft-syn   & 0.59 & 432 & 24 & 0 & 320 \\
                          &        & ft-h     & 0.96 & 724 & 24 & 0 & 28 \\
                          &        & ft-p     & 0.95 & 716 & 24 & 0 & 36 \\
                          &        & ft-hp    & 0.95 & 714 & 24 & 0 & 38 \\
                          &        & ft-all   & 0.96 & 721 & 24 & 0 & 38 \\
    \midrule
    SONIVA-Naming patient & Small  & Baseline & 0.41 & 126 & 87 & 1 & 310 \\
                          &        & ft-syn   & 0.52 & 188 & 87 & 1 & 248 \\
                          &        & ft-h     & 0.91 & 390 & 85 & 3 & 46 \\
                          &        & ft-p     & 0.90 & 383 & 86 & 2 & 53 \\
                          &        & ft-hp    & \textbf{0.92} & 397 & 86 & 2 & 39 \\
                          &        & ft-all   & 0.90 & 389 & 85 & 3 & 47 \\
    \cmidrule(l){2-8}
                          & Medium & Baseline & 0.45 & 150 & 88 & 0 & 286 \\
                          &        & ft-syn   & 0.48 & 165 & 87 & 1 & 271 \\
                          &        & ft-h     & 0.91 & 393 & 85 & 3 & 43 \\
                          &        & ft-p     & 0.91 & 389 & 86 & 2 & 47 \\
                          &        & ft-hp    & \textbf{0.92} & 397 & 86 & 2 & 39 \\
                          &        & ft-all   & 0.88 & 375 & 86 & 2 & 61 \\
    \bottomrule
  \end{tabular}
  }
  \label{accu}
\end{table}

\subsection{Accuracy Score Prediction}
The best-performing models for predicting the Semantic, Dysfluency, and Self-correction accuracy scores during the Naming task on healthy data were the Medium models initialized with \textit{ft-h} weights, achieving F1 Macro scores of 0.7449, 0.8390, and 0.7539, respectively (Table \ref{accu-pre}). For the Phonology metric, the best-performing model was the Small model also initialized with \textit{ft-h} weights, reaching an F1 Macro of 0.6424. Compared to their respective baselines, an average improvement of 7.24 ± 4.01\% was observed across all accuracy metrics. Although all models had moderate to high F1 Macro scores, there was a discrepancy between their performance on correct/unimpaired (class 1) and incorrect/impaired (class 0) trials. The F1 scores for class 1 trials were 0.9814, 0.9564, 0.9860, and 0.9744, whereas the F1 scores for class 0 trials were 0.5085, 0.7215, 0.5217, and 0.3103 for the Semantic, Dysfluency, Phonology, and Self-correction metric models, respectively.

In context of patient speech, the best-performing model for predicting Semantic accuracy scores was the Small model initialised with \textit{ft-p} encoder weights, reaching F1 Macro of 0.7659. For predicting Dysfluency and Self-correction, the best-performing models were the Small models initialized with baseline encoder weights, reaching F1 Macro of 0.9021 and 0.7112, respectively (Table \ref{accu-pre}). For Phonology, the best model was the Medium model initialized with \textit{ft-p} encoder weights, achieving F1 Macro of 0.6435. The observed improvement of fine-tuned models in Phonology and Semantics was in average 4.425 ± 0.175\%. In Dysfluency and Self-correction, the baseline models performed on average 1.45±0.06\% better than corresponding models initialized with \textit{ft-p} encoder weights. While the discrepancy between performance on class 1 and class 0 trials was still evident, it was lower than on healthy data. The F1 scores in class 1 trials were 0.9105, 0.9282, 0.8944 and 0.8604, whereas F1 scores in class 0 were 0.6213, 0.8759, 0.5279 and 0.4265 for Semantic, Dysfluency, Self-correction and Phonology metric models, respectively.

\begin{table}[ht]
  \centering
  \caption{F1 Macro for different accuracy metrics and testing datasets.}
  {\fontsize{8}{9.2}\selectfont
    \begin{tabular}{
    l!{\vrule width 0.3pt}
    l!{\vrule width 0.3pt}
    l!{\vrule width 0.3pt}
    c!{\vrule width 0.3pt}
    c!{\vrule width 0.3pt}
    c!{\vrule width 0.3pt}
    c
  }
    \toprule
    \textbf{Testing Dataset} & \textbf{Size} & \textbf{Encoder Weight} &
    \textbf{Semantic} & \textbf{Dysfluency} & \makecell{\textbf{Self-}\\\textbf{correction}} & \textbf{Phonology} \\
    \midrule
    SONIVA-Naming healthy & Small  & Baseline & 0.6196 & 0.7823 & 0.6124 & 0.5606 \\ 
                          &        & ft-h     & 0.6948 & 0.8375 & 0.6876 & \textbf{0.6424} \\
                          & Medium & Baseline & 0.7094 & 0.8009 & 0.6197 & 0.5334 \\
                          &        & ft-h     & \textbf{0.7449} & \textbf{0.8390} & \textbf{0.7539} & 0.6045 \\
    \midrule
    SONIVA-Naming patient & Small  & Baseline & 0.7234 & \textbf{0.9021} & \textbf{0.7112} & 0.5873 \\
                          &        & ft-p     & \textbf{0.7659} & 0.8870 & 0.6973 & 0.5771 \\
                          & Medium & Baseline & 0.7520 & 0.9017 & 0.6775 & 0.5975 \\
                          &        & ft-p     & 0.7370 & 0.8891 & 0.6613 & \textbf{0.6435} \\
    \bottomrule
  \end{tabular}
  }
  \label{accu-pre}
\end{table}

\subsection{Predictive Validity Analysis}
Self-correction accuracy scores were excluded from this analysis as only two patients were identified as impaired based on model predictions of this metric, making further statistical analysis impossible. 

The predictions of the best-performing accuracy prediction models were used to identify patients as impaired or unimpaired on each metric (Small Whisper initialized with \textit{ft-p} encoder weights, small Whisper initialized with baseline encoder weights, and Medium Whisper initialized with \textit{ft-p} encoder weights for Semantic, Dysfluency and Phonology, respectively). 

Patients identified as impaired based on model predictions had significantly lower speech fluency than those identified as unimpaired based on Semantic ($t(16)=-3.61, p=0.01$), Dysfluency ($t(16)=-2.3880, p=0.02$) and Phonology ($t(16)=-2.3880, p=0.02$) metrics. Significantly more impaired patients spoke English as a second language compared to unimpaired patients, based on Dysfluency and Phonology metrics ($p=0.02$). Additionally, a higher number of patients with previous stroke history were identified as impaired than unimpaired based on Semantic ($p=0.02$), Dysfluency ($p=0.04$) and Phonology ($p=0.04$) metrics.  The divergent validity analysis showed no significant differences between impaired and unimpaired patients for sex, LDL cholesterol or smoking status, as expected $(p>0.05)$ (See Appendix E).  

\section{Discussion}

The baseline Whisper model performed substantially worse on the verbatim transcription of single-word SONIVA-Naming healthy speech compared to previously reported results on healthy speech. 
Previous studies have reported WER as low as 3.40\% and 2.90\% for the baseline Whisper Small and Medium models, respectively \cite{Radford_2023}. However, these results were obtained using the LibriSpeech corpus \cite{panayotov2015librispeech}, a standard ASR benchmark consisting of continuous speech from audiobook recordings. In contrast, the markedly poorer performance observed in the current study likely reflects the short-form nature of the single-word Naming task-derived speech, which provides limited contextual information for decoding and therefore reduces transcription accuracy.

Although the synthetic speech dataset was free from natural speech variability and background noise, the baseline Whisper models also performed poorly on this dataset (WER = 92.85\% and 85.71\% for the Small and Medium models, respectively). The observed performance further supports that Whisper does not generalize directly to short-form speech and highlights the necessity of task-specific fine-tuning. Further, once fine-tuned, all models achieved their best performance on the synthetic dataset, as anticipated, since it represented the linguistically and acoustically simplest condition.

All fine-tuned Whisper models improved
transcription performance across datasets. Models trained on healthy and patient
data further improved the WER compared to those trained on
synthetic data, suggesting the importance of natural speech
diversity in improving transcription performance.

The fine-tuned model showed adequate performance for post-stroke speech transcription. In patients, the best performing model was the Medium \textit{ft-hp} model achieving a 26.35\% WER. However, this result was not significantly different from Medium \textit{ft-p} and \textit{ft-h} models (26.80\% WER ; $p>0.05$ and 27.72\% WER ; $p>0.05$, respectively). Similarly, target detection performance of Medium \textit{ft-hp}, \textit{ft-p} and \textit{ft-h} models was also comparable (0.92, 0.91 and 0.91 accuracy, respectively). 

This suggests that fine-tuning on age-matched, task-specific healthy speech may generalize sufficiently to patient data for the same task. This is particularly relevant in clinical applications since collection of healthy speech data is more accessible, and can facilitate a larger dataset for fine-tuning. 

Further, comparable performance between \textit{ft-h}, \textit{ft-p} and \textit{ft-hp} models on patient speech suggests that the main performance advantage comes from fine-tuning Whisper on short-form speech produced in the single-word Naming task, irrespective of the data source. 

The WER achieved by the \textit{ft-p} model was consistent with previous finding on wav2vec 2.0 and Whisper models fine-tuned on aphasic speech (WER 22.30\%–55.50\% depending on aphasia severity and 21.93\% WER, respectively) \cite{Torre_2021,Sanguedolce_2025}, and outperformed hyperparameter-tuned Whisper (WER 38.50\%) \cite{Sanguedolce_2023}. However, as prior work targeted continuous speech, direct comparison is limited.

Medium models achieved the best performance across most datasets but did not differ significantly in WER from the smaller variants. Given the higher computational cost and training time, Small models may offer a more efficient option for future clinical use, particularly when scaling to larger datasets.




Similar to the SONIVA-Naming test set, fine-tuned Whisper models showed markedly improved transcription on the unseen TORGO dataset compared to the baseline, underscoring the need to adapt pretrained models for out-of-domain clinical data. Nonetheless, the best-performing model (Medium \textit{ft-syn}) still produced a high WER of 73.49\%, well above clinical usability. Models fine-tuned on SONIVA patient data also performed poorly on TORGO (Medium WER: 74.62\%). This can be attributed to clinical heterogeneity: SONIVA targets stroke-induced aphasia involving both motoric and linguistic processing difficulties, whereas TORGO focuses on dysarthria and motoric impairments only. These results emphasize the importance of tailoring pretrained models to the specific linguistic and pathological features of each condition to achieve robust generalization and clinical utility.


In terms of target detection accuracy on the SONIVA-Naming task data, fine-tuned models showed higher True Positive and lower False Negative scores compared to baseline Whisper, improving clinical applicability. However, on patient data, the fine-tuned models produced more False Positive mistakes, as a result of failing to transcribe phonological errors made by patients (e.g., predicting target word ‘comb’ instead of ‘comg’). As Whisper is designed to predict the next most probable token in the sequence, these discrepancies are expected.

 Whisper was also used to complete a stroke-specific downstream task, aimed at estimating the level of speech impairment. In case of healthy data, the models initialized with \textit{ft-h} encoder weights had the best F1 Macro in accuracy score prediction, suggesting the importance of fine-tuning. However, for patient data, models initialized with \textit{ft-p} encoder weights improved performance on the Semantic and Phonology metrics but not on Dysfluency and Self-correction. Specifically, while the fine-tuned model outperformed previous studies that used Whisper encoder features for classifying dysfluencies in stuttering (F1 Macro of 0.71) \cite{Ameer_2024}, its performance was slightly lower than the baseline for Dysfluency. This reduction in performance suggests that while fine-tuning for verbatim transcription captures linguistic features like semantics and phonology, it may not capture more complex patterns in patient speech, such as hesitations and self-corrections, which are less vital
for transcription but crucial for Dysfluency and Self-correction predictions, leading to worse performance on these metrics for patient speech. It is, however, worth noting  that the absolute difference between the best baseline encoder models and their fine-tuned encoder versions was negligible (1.45±0.06\%).

In addition, the performance in the accuracy prediction task was higher on trials with unimpaired scores compared to impaired ones, especially in the healthy dataset. This is expected since in the case of healthy data, the models were exclusively fine-tuned on healthy speech where the number of impaired trials is inevitably limited, leading to smaller sample sizes and overall performance. 

Despite this difference between model performance on impaired and unimpaired trials, the patient -level clinical validity analysis confirmed that the predicted accuracy scores were clinically meaningful. A significant difference between patients identified as impaired or unimpaired based on model predictions was observed in speech fluency, English proficiency status, and previous stroke history, as expected \cite{Fonseca_2019,Hope_2015,Sharif_2022}. The results of the divergent validity analysis showed no differences in features unrelated to speech impairment such as cholesterol levels, sex and smoking status. 

\section{Limitations and Future Work}

We note several limitations and directions for future work. First, patients were not stratified by speech impairment severity when splitting into training, validation, and test sets, which may have led to uneven distributions and limited generalizability across impairment levels. Nonetheless, this should not affect the main conclusions about Whisper, as similar findings were observed in healthy data.


In the accuracy prediction task, performance could be improved by addressing class imbalance through multiple binary classification tasks (instead of a multi-class one) and by using approaches alternative to downsampling. For example, data perturbation, modifying audio features such as pitch, frequency, or tempo, to create more samples has been used in disordered and aphasic speech transcription \cite{Geng_2020, moell2022speech, yuan-etal-2022-data}. Alternatively, in-domain data augmentation, such as incorporating similar tasks from datasets like AphasiaBank \cite{Macwhinney_2011}, could also be applied \cite{gale-etal-2022-post}.

Whisper’s transcription accuracy could also be evaluated using beam search instead of greedy decoding, to create a more complete assessment of its applicability to stroke-specific data. Greedy decoding selects the most probable token at each step in the output sequence, whereas beam search considers multiple candidate sequences and chooses the final output based on cumulative probability \cite{Radford_2023,chen2018stableeffectivelearningstrategy}.


In order to obtain a broader overview on the application of foundation models to pathological speech in stroke, Whisper should be compared with other transformer-based architectures, such as Seamless M4T v2, a multimodal model capable of speech transcription that has shown competitive results compared to Whisper on ASR tasks \cite{Barrault_2023}. Finally, future work should explore whether token-level loss analyses, encoder probing, or modality-specific pretraining can help disentangle the extent to which current models capture generalizable acoustic markers of clinical speech, particularly in settings where contextual compensation from the decoder is limited, such as single-word utterances.

\section{Conclusion}
We evaluated the applicability of Whisper, a general foundation model, to a specific clinical case study on a single-word speech task in patients with stroke. Our findings confirmed that additional fine-tuning on single-word data is required to achieve efficient transcription performance that can facilitate clinically useful downstream tasks. Further, comparisons of fine-tuned model performance on in-domain and out-of-domain neurological disorder datasets suggests further need for model adaptation to disorder-specific data. 
With such fine-tuning we can advance the automated analysis of pathological speech allowing for more efficient diagnosis and monitoring of stroke.



\appendix

\newpage
\section{Sample demographics of the SONIVA-Naming dataset}
\begin{table}[ht]
  \centering
  \caption{Sample demographics of the SONIVA-Naming dataset.}
  {\fontsize{8}{9.2}\selectfont
  \begin{tabular}{
    l!{\vrule width 0.3pt}
    c!{\vrule width 0.3pt}
    c
  }
    \toprule
    \textbf{Demographic Feature} & \textbf{SONIVA-Naming healthy} & \textbf{SONIVA-Naming patients} \\
    \midrule
    \multicolumn{1}{l!{\vrule width 0.3pt}}{} & \small N = 132 & \small N = 87 \\
    \multicolumn{1}{l!{\vrule width 0.3pt}}{} & \multicolumn{2}{c}{Mean (Standard Deviation)} \\
    \midrule
    Age & 61.6 (10.8) & 61.8 (13.6) \\
    Sex$^{***}$ &  &  \\
    \hspace{0.5cm}Male:Female & 52:77 & 62:25 \\
    \hspace{0.5cm}Missing data & 3 & 0 \\
    English Language$^{*}$ &  &  \\
    \hspace{0.5cm}Non-native & 27 & 30 \\
    \hspace{0.5cm}Native & 102 & 57 \\
    Education level$^{***}$ &  &  \\
    \hspace{0.5cm}School & 36 & 45 \\
    \hspace{0.5cm}Degree & 43 & 37 \\
    \hspace{0.5cm}Post-graduate & 37 & 3 \\
    \hspace{0.5cm}Missing data & 16 & 2 \\
    \bottomrule
  \end{tabular}
  \vspace{0.3em}
  \\
  {\footnotesize $^{*}p<0.05$;\quad $^{***}p<0.001$}
  }
  \label{tab:demographics}
\end{table}

 \section{Whisper Architecture}

Whisper has an encoder-decoder Transformer architecture \cite{Radford_2023} with different model size availability, corresponding to the number of parameters (39M-1550M) and encoder-decoder block layers (4-32). Each encoder block of Whisper consists of a self-attention layer and a multilayer perceptron. The model processes audio input in a form of an 80-channel log-Mel spectrogram, passing it through two initial convolutional layers followed by the encoder blocks. The encoder outputs information-dense context vectors, which are a high-level feature representation of the input audio sequence, encompassing acoustic, positional, and contextual information through self-attention mechanisms and positional embedding. 

The decoder blocks mirror the architecture of the encoder blocks, incorporating an additional cross-attention layer, which enables the decoder to focus on the outputs of the encoder. This architecture allows the decoder to process the context vectors produced by the encoder and the tokenized input text sequences. Whisper utilizes a Byte-Pair Encoding tokenizer to break down the input text into smaller sub-word units \cite{sennrich-etal-2016-neural}. During training, tokenized text sequences correspond to the audio input's ground truth labels, while during inference, they consist of previously predicted tokens. Additionally, Whisper utilizes a range of special tokens, which signal the start and end of the sequence, task type and language. The decoder's raw output is passed through a final softmax layer to obtain the next token prediction. 

To complete the accuracy classification task,
WhisperforAudioClassification model configuration \cite{WhisperForAudioClassification} was used, which retains the same Whisper encoder architecture, but replaces the decoder with a linear classifier projection head, consisting of two linear layers.

\newpage
\section{The general study pipeline}
\begin{figure}[h!]
    \centering
    \includegraphics[width=0.8\linewidth]{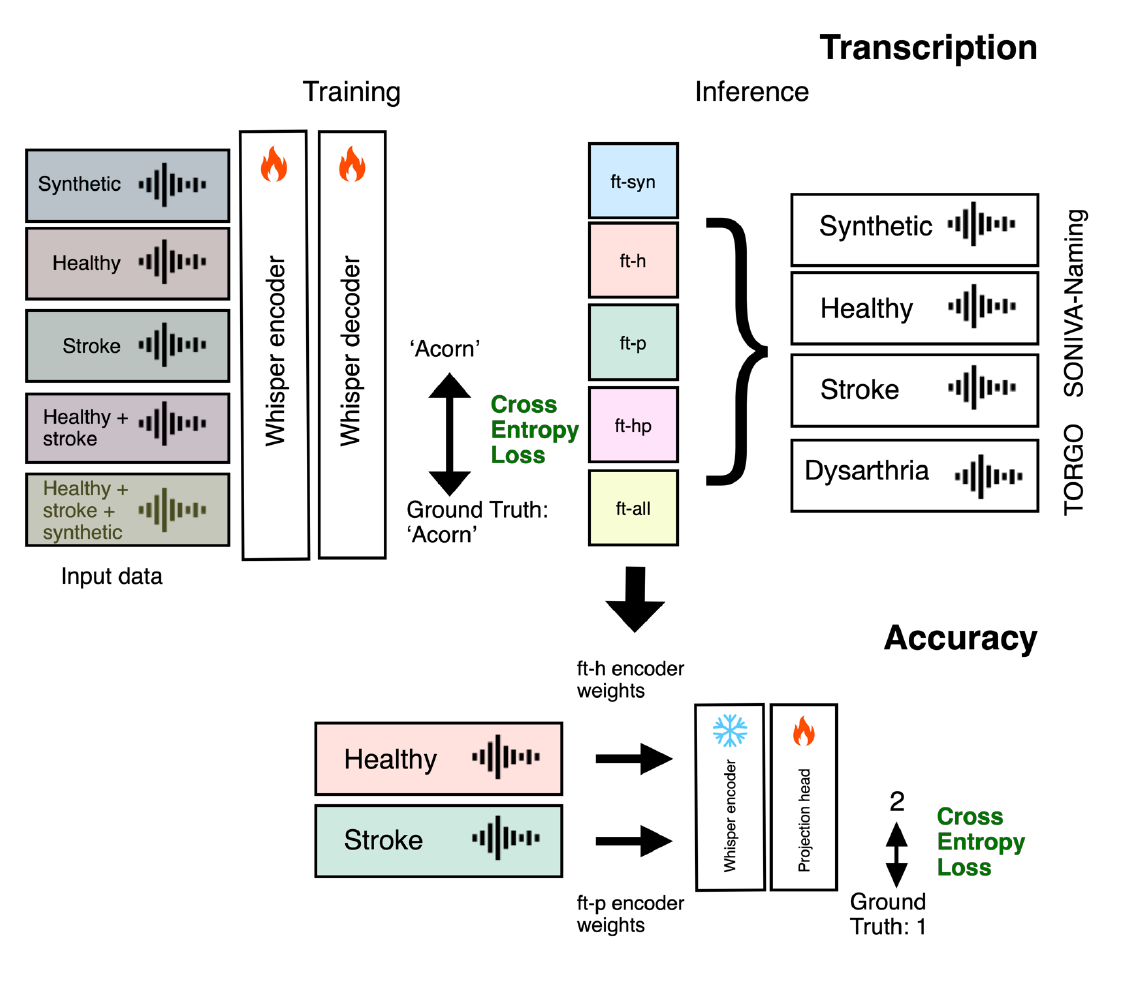}
    \caption{General study pipeline. Overview of the training and inference pipeline for verbatim transcription and accuracy prediction models.
Transcription: Training – Whisper models were fine-tuned on speech data from 5 different datasets. A synthetic dataset of audio-transcription word pairs based on stimuli in the Naming task, SONIVA-Naming healthy dataset, collected from healthy age-matched participants performing the Naming task, SONIVA-Naming patient dataset, collected from patients with stroke performing the Naming task, a combined dataset of SONIVA-Naming healthy and patient data, and a combined dataset of all available data (synthetic, SONIVA-Naming healthy and SONIVA-Naming patient). All models were trained using cross-entropy loss to transcribe spoken words (e.g. “acorn”) from the Naming task input audio. Inference – Fine-tuned models (\textit{ft-syn}, \textit{ft-h}, \textit{ft-p}, \textit{ft-hp} and \textit{ft-all}) were evaluated on 4 separate datasets – synthetic, SONIVA-Naming healthy, SONIVA-Naming patient, and an unseen testing dataset of dysarthric speech derived from the TORGO database.
Accuracy: Training – accuracy prediction models were trained on SONIVA-Naming derived healthy and patient speech in a linear probing framework. The encoder layer of the models was frozen and the classification projection head was trained to predict accuracy scores using cross-entropy loss (e.g “2"). The encoder weights of the models were initialized at baseline or with the weights derived from the models trained for verbatim transcription of healthy or patient speech  }
    \label{fig:soniva}
\end{figure}

\section{Additional training details}
All model training was conducted on a single NVIDIA RTX 6000 GPU. Mixed-precision arithmetic (fp16) and gradient checkpointing was implemented to improve computational efficiency and optimise memory usage. Model training and evaluation were conducted using the PyTorch framework \cite{paszke2019pytorch} and HuggingFace Transformers library \cite{wolf2019huggingface}. 

\newpage
\section{Predictive Validity Analysis}
\begin{figure}[h]
    \centering
    \includegraphics[width=0.8\linewidth]{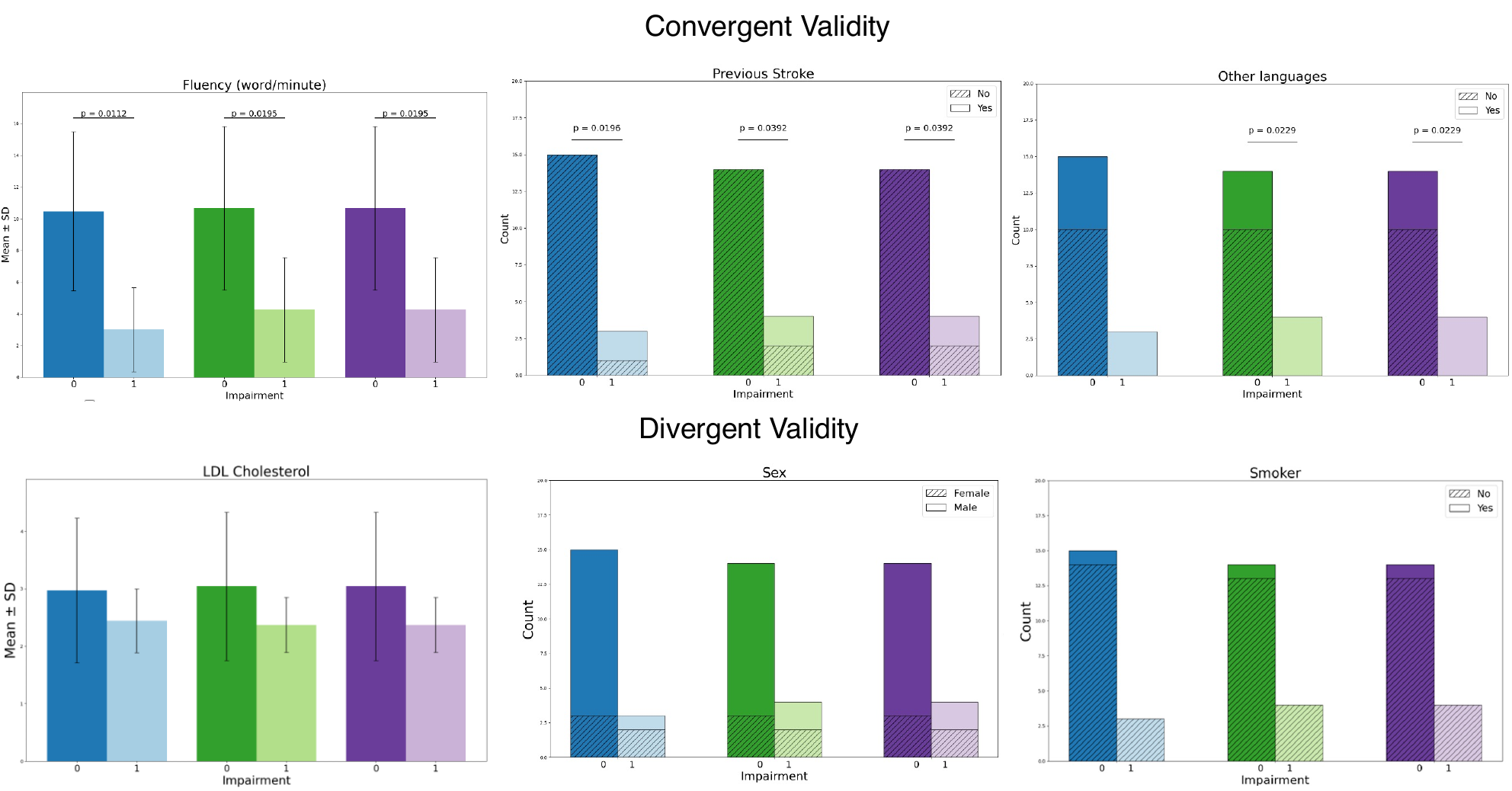}
    \label{plot_validity}
    \caption{Characteristics of patients identified as impaired (0) or unimpaired (1) from the SONIVA-Naming patient dataset. Predicted impairment status is
displayed separately for Semantic (blue), Dysfluency (green), and Phonology (purple) metrics. Speech fluency, previous stroke history, English as second
language (other languages), LDL cholesterol level, sex and smoking status were assessed. Error bars indicate standard deviation. Between group significance
is denoted with p-values (uncorrected).}
    \label{plot_validity}
\end{figure}

\end{document}